\documentclass[12pt]{iopart}
\usepackage{iopams}
\usepackage{indentfirst}
\usepackage{braket}
\usepackage{makeidx}
\usepackage{slashed}
\usepackage{graphicx}
\usepackage{xcolor}
\usepackage{hyperref, comment}
\usepackage[normalem]{ulem}

\begin{document}
\title{Stochastic thermodynamics of inertial-like Stuart-Landau dimer}
\author{Jung-Wan Ryu$^{1,2}$, Alexandre Lazarescu$^3$, Rahul Marathe$^{4}$, and Juzar Thingna$^{1,2}$}
\address{%
$^1$Center for Theoretical Physics of Complex Systems, Institute for Basic Science (IBS), Daejeon 34126, Republic of Korea.}
\address{%
$^2$ Basic Science Program, Korea University of Science and Technology (UST), Daejeon 34113, Republic of Korea.}
\address{%
$^3$ Institut de Recherche en Math\'ematiques et Physique, UCLouvain, Louvain-la-Neuve, Belgium.}
\address{%
$^4$ Department of Physics, Indian Institute of Technology, Delhi, Hauz Khas 110016, New Delhi, India.}
\ead{\mailto{jungwanryu@ibs.re.kr}; \mailto{alexandre.lazarescu@uclouvain.be};
\mailto{maratherahul@physics.iitd.ac.in};
\mailto{jythingna@ibs.re.kr}.}
\begin{abstract}
Stuart-Landau limit-cycle oscillators are a paradigm in the study of coherent and incoherent limit cycles. In this work, we generalize the standard Stuart-Landau dimer model to include effects due to an inertia-like term and noise and study its dynamics and stochastic thermodynamics. In the absence of noise (zero-temperature limit), the dynamics show the emergence of a new bistable phase where coherent and incoherent limit cycles coexist. At finite temperatures, we develop a stochastic thermodynamic framework based on the dynamics of a charged particle in a magnetic field to identify physically meaningful heat and work. The stochastic system no longer exhibits the bistable phase but the thermodynamic observables, such as work, exhibit bistability in the temporally metastable regime. We demonstrate that the inertial-like Stuart-Landau dimer operates like a machine, reliably outputting the most work when the oscillators coherently synchronize and unreliable with minimum work output when the oscillators are incoherent. Overall, our results show the importance of coherent synchronization within the working substance in the operation of a thermal machine.
\end{abstract}
\maketitle

\section{Introduction}

Macroscopic heat engines that convert wasteful heat into useful mechanical work have been a cornerstone in establishing the laws of thermodynamics \cite{Muller07}. Due to the technological advancements over the past several decades, there is a renewed interest in meso- and micro-scopic engines~\cite{Bechinger10,QuintoSu14,Martinez16,Martinez17}. In these nanoscale machines, which are connected to macroscopic heat baths, thermal fluctuations that randomize the dynamics of the small working substance are no longer negligible \cite{Seifert08}. Such small systems thus, can not be described by traditional thermodynamics and has led to the advent of stochastic thermodynamics \cite{Sekimoto97,Seifert12,Murashita16} that provides a rigorous mathematical framework to define not only the average thermodynamic quantities like heat and work, but also their distributions.

Nanoscale heat engines have been extensively studied exploring the roles of external time dependent driving \cite{Marathe07, Marathe17}, active dissipation~\cite{Marathe18,Marathe19,Marathe20,Lee20}, particle-particle interaction~\cite{Benenti20}, network topology~\cite{Ashida21}, collective phenomenon~\cite{Herpich18, ZhangNatPhys19,Hong20}, etc., to improve the performance metrics of the engine. We focus on the fascinating phenomenon of synchronization~\cite{Pik01} and explore whether the working substance comprising of synchronizing particles can make better machines? We consider the paradigmatic model, Stuart-Landau~\cite{Lan44,Stu60} dimer, used for understanding collective behaviours such as synchronisation and oscillation suppression in coupled nonlinear oscillators \cite{Kur84,Aoy95,Sax12,Mat90,Kos13}, and extend this overdamped-deterministic model to the underdamped-stochastic regime, incorporating the effects of inertia and temperature. Surprisingly, even though the inertial-like stochastic Stuart-Landau dimer has never been explored, its famous counterpart the Kuramoto model has been extensively studied. The inertial stochastic Kuramoto model is used to describe synchronisation of fireflies \cite{Tan97a,Tan97b}, disordered Josephson Junction arrays \cite{Tre05}, power grids \cite{Doe13}, and has also been used to explore nonequilibrium phase transitions~\cite{Gupta14,Gupta14Jstat}.

In this work, our first main aim is to understand the effects of the inertial-like term on the phase diagram for the nonlinear oscillator model and explore how synchronisation affects the work and heat exchanges which characterise it as a thermodynamic machine. Similar to the inertial Kuramoto model~\cite{Bel16,Yua17}, we find that due to the inertia-like term the zero-temperature phase diagram develops bistability leading to a coexistence of coherent and incoherent limit cycles. Secondly, we build a stochastic thermodynamics framework and show that the thermodynamic observables (heat and work) depend strongly on the physical interpretation of the dynamical equations being analyzed. In other words, given the dynamical equations there does not exist a \emph{unique} mathematical framework to obtain physically meaningful heat and work. Specifically, the definitions of thermodynamic quantities depend on how one segregates the various parameters (system or reservoir parameters) appearing in the dynamical equations and if the dynamics has the correct time-reversal symmetry~\cite{Grmela97, Grmela97II, Kraaij18}. The physically meaningful thermodynamics depends also on whether the system is described by underdamped or overdamped dynamics \cite{Maes00}, whether in a Newtonian or accelerating frame \cite{Horowitz16}, and whether any of the forces are pseudovectors (such as magnetic fields) \cite{Barbier19}. 

The paper is organised as follows: In Sec.~\ref{sec:SLD}, we start with defining the stochastic inertial-like Stuart-Landau dimer, and we explore its deterministic phase diagram. This will guide us while studying the stochastic version of the model by the introducing the thermal noise. In Sec.~\ref{sec:ST}, we show that, though the inertial-like Stuart-Landau oscillator might appear to be driven by non-conservative environmental forces (involving work production even without dimer interaction), it can in fact be seen as an equilibrium charged particle in a magnetic field, described in a rotating reference frame; this gives us a well-motivated physical frame for defining work and heat in the interacting case. In Sec.~\ref{sec:Num}, we numerically study the behavior of our model in terms of average work in different bistable regimes, for this we take a cue from the deterministic case. We also study the full probability distributions and the reliability of our system. We conclude in Sec.~\ref{sec:Con}.  

\section{Stuart-Landau Dimer}\label{sec:SLD}
The Stuart-Landau oscillator (SLO) is a prototypical system exhibiting Hopf bifurcation and limit cycle oscillations that can reveal universal features of many practical systems \cite{Lan44, Stu60, Kur84, Aoy95}. In this section, we introduce and study the inertial-like Stuart-Landau dimer, comprising of two coupled SLOs, and investigate the effects of the inertial-like term on its phase diagram. The specific parameters entering the equations will be physically justified in the next section.

\subsection{Inertial-like Stuart-Landau dimer}
Our starting point is the system of four coupled underdamped Langevin equations that govern the dynamics of each pair of coordinates $(x_i,y_i)$ of the oscillators:
\begin{eqnarray} 
\label{xyLangevin}
m \ddot{x}_1 &=& (R-x_1^2-y_1^2)~x_1-\gamma\omega_1 y_1-k~(x_1-x_2)-\gamma\dot{x}_1 + \xi_1^R, \label{Lx1}\\
m \ddot{y}_1 &=& (R-x_1^2-y_1^2)~y_1+\gamma\omega_1 x_1-k~(y_1-y_2)-\gamma\dot{y}_1 + \xi_1^I, \label{Ly1}\\
m \ddot{x}_2 &=& (R-x_2^2-y_2^2)~x_2-\gamma\omega_2 y_2-k~(x_2-x_1)-\gamma\dot{x}_2 + \xi_2^R, \label{Lx2}\\
m \ddot{y}_2 &=& (R-x_2^2-y_2^2)~y_2+\gamma\omega_2 x_2-k~(y_2-y_1)-\gamma\dot{y}_2 + \xi_2^I. \label{Ly2}
\end{eqnarray}
Both oscillators have the same mass $m$. The parameter $R$ represents the strength of the external potential, and acts as a control parameter such that in the overdamped-uncoupled-noiseless limit ($\gamma >> m$, $k\rightarrow 0$, and $\xi_l^{R(I)}(t) = 0$) the oscillators settle in their own limit cycle with radius $\sqrt{R}$ for $R >0$ and to a fixed point if $R < 0$. Each two-dimensional oscillator is governed by a frequency $\omega_l$ (scaled by the damping coefficient) and the oscillators are coupled via a spring with strength $k$. The parameter $\gamma$ represents the phenomenological Stokesian damping coefficient. The random noise $\xi_j^{R,I}$, $j=1,2$ are Gaussian with zero mean and $\langle \xi_l^{\alpha}(t)*\xi_j^{\beta}(t') \rangle = 2\gamma_j k_{B} T_j\delta(t-t') \delta_{\alpha,\beta}\delta_{l,j}$, where $\langle\cdot\rangle$ represents the ensemble average over the noises. The strength of the noise is chosen such that the fluctuation dissipation relation holds. Throughout this work we will set $R=1$ as the unit scale in which other parameters are measured, $\omega_{1} = 1$ and $\omega_{2} = \omega_{1} + \Delta \omega$, unless mentioned otherwise. The temperatures $T_1$ and $T_2$ could be different, but we will focus on the case where they are the same, so that the nonequilibrium driving comes only from the deterministic part of the dynamics.

A potentially useful and equivalent way to rewrite these equations is to use the complex notation $z_1=x_1+i y_1, z_2=x_2+i y_2$, which yields the following compact expressions:
\begin{eqnarray}
\label{ine_noise}
    m \ddot{z}_1 &=& \left(R + i \gamma\omega_1 - |z_1|^2\right) z_1 - \gamma \dot{z}_1 + k ( z_2 - z_1 ) + \xi_1, \nonumber\\
    m \ddot{z}_2 &=& \left(R + i \gamma\omega_2 - |z_2|^2\right) z_2 - \gamma \dot{z}_2 + k ( z_1 - z_2 ) + \xi_2. 
\end{eqnarray}
Above the complex noise $\xi_l = \xi_l^R + i \xi^I_l$. The above set of equations with complex variable $z_l$ converges to the standard Stuart-Landau dimer in the overdamped limit $\gamma >> m$ and the underdamped version will hereon be referred to as the inertial-like Stuart-Landau Dimer (ISLD). Note that we consider the coordinates $(x_i,y_i)$ of the oscillator to be independent of each other and to describe position coordinates, due to which the presence of the second-order term refers to an inertia-like term. The independence between $x$ and $y$ is not necessary \cite{Karakaya2019}, but it is true in several systems of interest, e.g. neuron firing experiments wherein for a single SLO there are two independent variables namely the amplitude of the pulse and the duration between two firings \cite{Aoy95} or Nano-Electro Mechanical Systems (NEMS) wherein if one goes beyond  the slow variation of the amplitude approximation then a second order derivative for the complex variable $z$ appears naturally~\cite{Matheny2019,Lifshitz09}. In the case of $x$ and $y$ being canonically conjugate (in the sense of Hamiltonian dynamics) the standard Stuart-Landau model itself would include an inertial term and the complex variable equations can be dealt with the use of Wirtinger derivatives. Throughout this work, we do not consider this case and refer the interested readers to Refs.~\cite{Borlenghi17, Borlenghi18}. Since even the standard Stuart-Landau equation could include effects of inertia, we use the term inertia-\emph{like} to avoid conceptual pitfalls.

%one may sometimes use the Wirtinger derivatives to deal with such complex variable equations, see e.g. Refs.~\cite{Borlenghi17, Borlenghi18}, but in those cases the real ($x$) and imaginary ($y$) components of the complex variables are required to be canonically conjugate with each other (in the sense of Hamiltonian dynamics), which is not the case here.

To the best of our knowledge, the noisy inertial-like Stuart-Landau dimer has never been explored before. In the remainder of this section we will fully explore the phase space and bifurcation diagram of this model in the zero temperature limit to illustrate the existence of bi-stability that arises only in the underdamped regime.

\subsection{Phase diagram and temporal behavior}
\begin{figure}
\begin{center}
\includegraphics[width=\columnwidth]{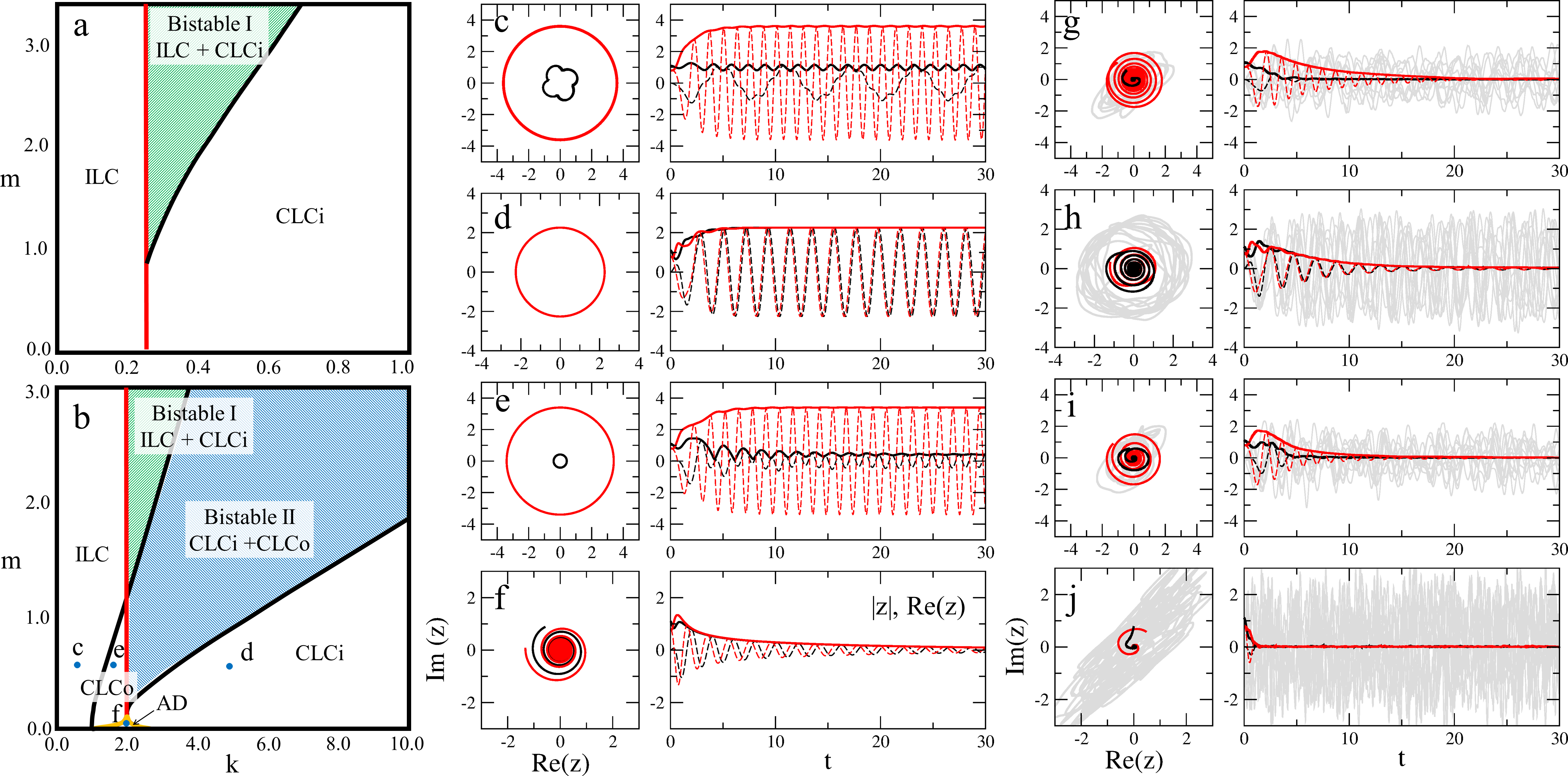}
\caption{Phase diagrams when $\Delta \omega = \omega_2-\omega_1 = 0.5$ (panel {\bf a}) and $\Delta \omega = 4.0$ (panel {\bf b}). ILC, CLCi, CLCo, and AD represent incoherent limit cycle, in-phase coherent limit cycle, out of phase coherent limit cycle, and amplitude death. The coexistence of incoherent and coherent limit cycles is shaded in green (bistable I) whereas the coexistence of different types of coherent limit cycles is in blue (bistable II). Trajectories on the complex plane and time series of the two oscillators when ($k$, $m$) = (0.5, 0.5) (panels {\bf c} and {\bf g}: ILC), (5.0, 0.5) (panels {\bf d} and {\bf h}: CLCi), (1.5, 0.5) (panels {\bf e} and {\bf i}: CLCo), and (2.0, 0.1) (panels {\bf f} and {\bf j}: AD). Panels ({\bf c}-{\bf f}) are the noiseless limit with black and red colors representing the first and second oscillator, respectively. The real parts $\mathrm{Re}(z)$ are the dashed lines in the temporal plots and the amplitudes $|z|$ are represented by the solid lines. Panels ({\bf g}-{\bf j}) are the noisy oscillators with grey color being 10 selected trajectories of $\mathrm{Re}(z_{1})$ of the first oscillator. The dashed black (red) ($\left<\mathrm{Re}(z_{1})\right>$) and solid black (red) ($|\left<z_{1}\right>|$) represent trajectories of the first (second) oscillator averaged over 1000 realizations. Throughout we have set $\gamma=1$ and the points marked {\bf c}, {\bf d}, {\bf e}, and {\bf f} in blue in panel {\bf b} correspond to parameter values used in panels {\bf c}, {\bf d}, {\bf e}, and {\bf f}, respectively. The orange and red boundaries in panels {\bf a} and {\bf b} are determined analytically from linear stability analysis as shown in \ref{append:LSA}. The black boundaries are numerically obtained using the bifurcation diagrams similar to Fig.~\ref{fig3}. Temperature used in panels {\bf g, h, i} is $K_BT_L=k_BT_R=4.0$ and in {\bf j} is $k_BT_L=k_BT_R=20.0$.}
\label{fig2}
\end{center}
\end{figure}

The physical steady-state phases of the ISLD are classified in Figs.~\ref{fig2} {\bf a} and {\bf b} for the noiseless limit with small and large frequency difference $\Delta \omega$, respectively. We first focus on the noiseless case that exhibits a complex phase diagram and then explain the effect of noise. For simplicity, we have set $\gamma=1$.\\
~\\
\noindent {\it Noiseless without an inertia-like term.--} In the case without an inertia-like term ($m=0$), there are three distinct phases, an incoherent limit cycle (ILC), amplitude death (AD), and an in-phase coherent limit cycle (CLCi). For $\gamma\Delta \omega < 1$ the AD regime vanishes and we have an exceptional point that demarcate the ILC to CLCi transition. For $\gamma\Delta \omega > 1$ the AD region emerges such that its area in the $k$-$\gamma\Delta\omega$ plane increases~\cite{Erm90, Ryu15}. In this overdamped case, the boundaries of the AD regime can be analytically predicted using linear stability analysis and for $\gamma\Delta \omega < 1$ the critical coupling $k = k_{c} = \gamma\Delta \omega/2$~\cite{Ryu15} (exceptional point) segregates the ILC and CLCi regions. 

The physical picture in this regime is quite intuitive, for small $k$ each oscillator vibrates with its own frequency and hence there is no notion of synchronization giving an ILC due to the absence of noise ($T_l=0$), Fig.~\ref{fig2} {\bf c}. Note here that the second oscillator (red solid line) has a very small oscillation in its amplitude not visible due to the scale implying that it is indeed an ILC oscillator. As $k$ increases, the frequency mismatch between the oscillators decreases and the oscillators motion destructively interfere to give rise to a fixed point solution known as AD, Fig.~\ref{fig2} {\bf f}. For large values of $k$, since there is a strong coupling between the oscillators a self-organizing mechanism kicks in and the oscillators synchronize with their frequencies and amplitudes becoming equal with a finite phase difference as shown in Fig.~\ref{fig2} {\bf d}. This is known as the in-phase coherent limit cycle (CLCi) regime since the phase differences are smaller than $\pi/2$. As the $k$ increases further, the phase difference decreases and in the limit $k\rightarrow \infty$ the oscillators are perfectly in phase.\\
~\\
\noindent {\it Noiseless with an inertia-like term: small frequency difference $\gamma\Delta \omega < 1$.--}
Adding an inertia-like term to the system leads to rich limit cycle dynamics. If $m$ is sufficiently large with $\gamma\Delta \omega$ small, incoherent (ILC) and coherent (CLCi) limit cycles can coexist as seen in Fig.~\ref{fig2} {\bf a}. Thus, the ISLD exhibits bistability, similar to that of Kuramoto model with inertia \cite{Ji13}. Such bistable behavior is also known to exist in dimers exhibiting explosive synchronization for large complex networks~\cite{Gardenes2011,Leyva2012}. The boundaries of various phases are no longer analytically tractable unlike the $m=0$ case. The boundaries that we can obtain from linear stability is one that separates the AD region (orange line in Fig.~\ref{fig2} {\bf b}) and the red line in Figs.~\ref{fig2} {\bf a} and {\bf b} that corresponds to an exceptional `line' (see \ref{append:LSA} for more details).

\begin{figure}
\begin{center}
\includegraphics[width=\columnwidth]{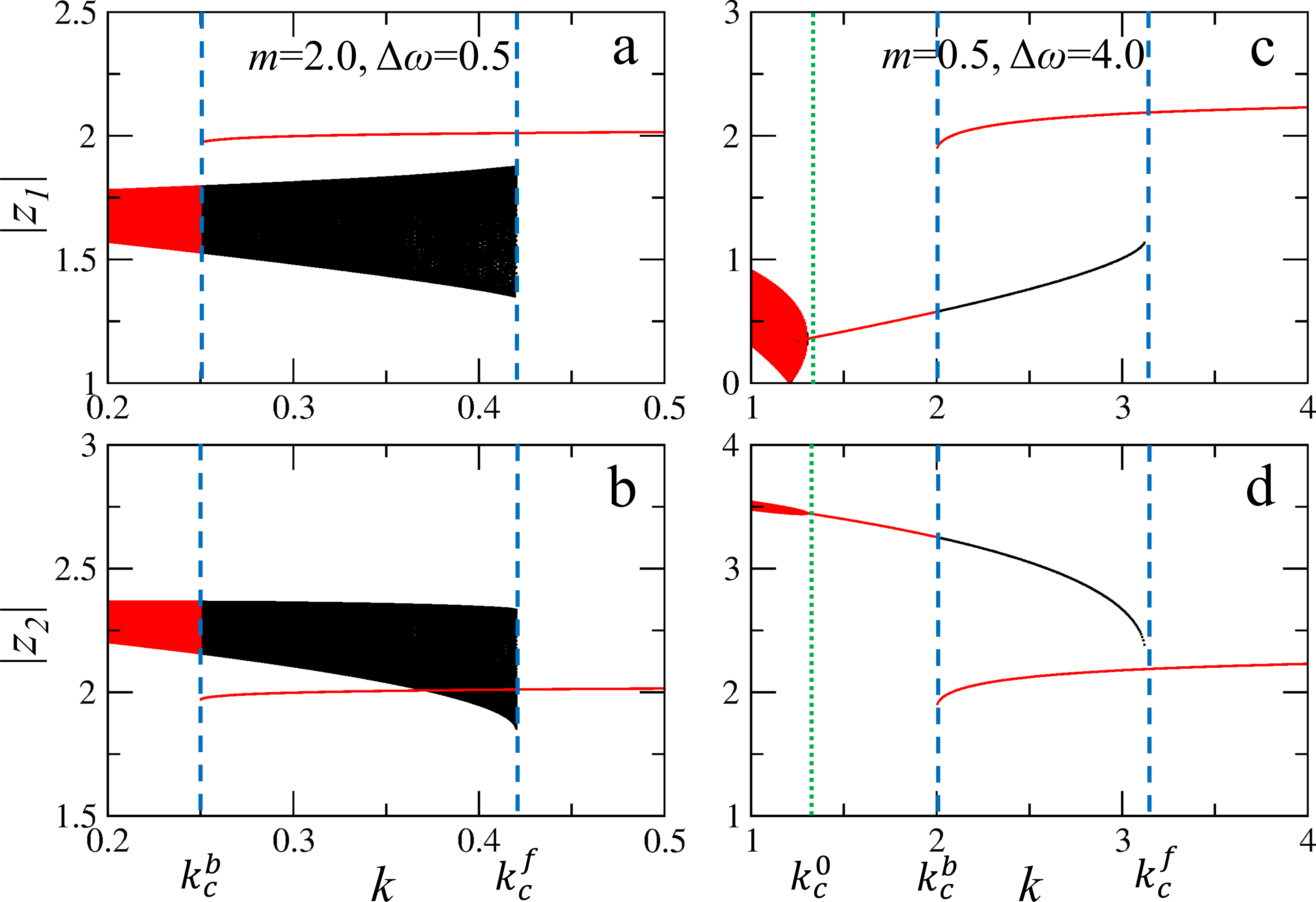}
\caption{Bifurcation diagram of amplitudes, $|z_{1}|$ and $|z_{2}|$, of two oscillators when $(m, \Delta \omega) = (2.0, 0.5)$ (panels {\bf a} and {\bf b}) and $(m, \Delta \omega) = (0.5, 4.0)$ (panels {\bf c} and {\bf d}). Black and red colors denote forward and backward directions of $k$ (see text for more details). The black and red spreads represent incoherent oscillator regions (ILC), while lines emerge for coherent oscillators (CLCi and CLCo). Blue dashed lines represent backward and forward transition points, $k_{c}^{b}$ and $k_{c}^{f}$. Green dotted lines represent the transition points between ILC and CLCo, $k_{c}^{0}$.}
\label{fig3}
\end{center}
\end{figure}

Figures~\ref{fig3} {\bf a} and {\bf b} show the bifurcation diagrams of amplitudes, $|z_i|$, of two oscillators when $m=2.0$ and $\gamma\Delta \omega = 0.5$ as a function of $k$. In order to obtain the bifurcation diagram, we initialize the system in a state given by the asymptotic limit of the previous coupling $k-\delta k$ plus a small perturbation (forward direction, black lines in Fig.~\ref{fig3}). In the ILC regime ($k < k_c^f$), the amplitude oscillates as seen by the spread of $|z_i|$ (black region hidden under red in Figs.~\ref{fig3} {\bf a} and {\bf b}). At the critical value of the coupling $k_c^f$ a small fluctuation of initial conditions triggers a discontinuous jump and the oscillators synchronize in phase (black line, hidden under red, in Figs.~\ref{fig3} {\bf a} and {\bf b}). In the reverse direction~\cite{note1}, (red lines in Fig.~\ref{fig3}) for $k > k_c^b$ the oscillators remain sychronized in phase (CLCi) and we observe a single red line in Figs.~\ref{fig3} {\bf a} and {\bf b}. At the backward critical value of the coupling $k_c^b$ the system transits from CLCi to an ILC which is maintained for all $k < k_c^b$. Thus, the bifurcation diagrams not only allow us to obtain all the phase boundaries numerically, e.g., the boundary (black solid line) separating the Bistable I and CLCi phase in Fig.~\ref{fig2} {\bf a}, but also clearly shows the hysteric synchrony of incoherent and coherent states in the region between $k_{c}^{b}$ and $k_{c}^{f}$~\cite{Tan97a, Tan97b} that crucially depends on the choice of initial conditions indicating bistability.\\
~\\
{\it Noiseless with an inertia-like term: large frequency difference $\gamma\Delta \omega > 1$.--}
When the difference of frequencies of the two oscillators is large the dynamics becomes highly complex in the presence of an inertia-like term as compared to the small $\gamma\Delta\omega$ regime. In this parameter regime, there exists an out of phase coherent limit cycle (CLCo) in which both oscillators have the same frequency but their amplitudes are different and they differ by a phase larger than $\pi/2$ (see Fig.~\ref{fig2} {\bf e}). At small mass $m$, unlike the low frequency difference case, a new boundary emerges (solid black line in Fig.~\ref{fig2}~{\bf b}) that separates the ILC and CLCo regime. For large $m$, the boundary at the critical coupling $k=k_c=\gamma\Delta\omega/2$ re-emerges and separates either the CLCo and bistable or ILC and bistable regime. The large frequency difference regime also produces a new bistable regime in which two different types of coherent states can coexist, CLCi and CLCo (see blue region in Fig.~\ref{fig2} {\bf b}). This type-II bistable region is distinctly different that the type-I, encountered in the small frequency difference regime that permits the coexistence of ILC and CLCi. At large $m$, both type-I and type-II bistable regions exist at large frequency differences, giving a rich complex phase diagram.

Figures~\ref{fig3} {\bf c} and {\bf d} show the bifurcation diagrams of amplitudes, $|z_i|$, of two oscillators when $m=0.5$ and $\gamma\Delta \omega= 4.0$ (large frequency difference) as a function of $k$. As $k$ increases (black line hidden under red), the amplitude $|z_{i}|$ changes from the ILC to CLCo at a critical value $k_{c}^{0}$ and hence we see the spread reduce to a single line around $k\approx 1.3$. As $k$ increases further, the oscillators are synchronized (CLCo) up to a $k = k_{c}^{f}$. At this point, a small fluctuation of initial conditions triggers a discontinuous jump to the CLCi regime. In the reverse direction when $k$ is reduced (red lines in Figs.~\ref{fig3} {\bf c} and {\bf d}), the oscillators maintain an in-phase synchronous behaviour (CLCi) up to a different $k = k_{c}^{b}$ at which the system transits to being out-of-phase (CLCo). At $k_{c}^{0}$ the system returns to the ILC phase which shows a spread in the bifurcation diagram. In this case again the bifurcation diagram shows a hysteresis and allows us to obtain the boundaries separating the CLCo phase (black lines in Fig.~\ref{fig2} {\bf b}). In the bistable regime, CLCi and CLCo can be distinguised based on the amplitude values of the oscillators. For CLCi, both oscillators have the same amplitude (compare red lines in Figs.~\ref{fig3} {\bf c} and {\bf d}) whereas for CLCo they have different amplitudes (compare black lines in Figs.~\ref{fig3} {\bf c} and {\bf d}).\\
~\\
{\it Noiseless CLC solution.--} For the system with an inertia-like term, we can find the semi-analytic solution in the CLC regime assuming that the two oscillators have the same frequency but different amplitude and phase. Thus, using $z_1=r_1e^{i(\Omega t + \phi_1)}$ and $z_2=r_2e^{i(\Omega t + \phi_2)}$ in Eq.~(\ref{ine_noise}) with $\xi_1(t)=\xi_2(t)=0$ and solving for $r_i$, $\Omega$, and $\Delta\phi=\phi_1-\phi_2$ we obtain,
\begin{eqnarray}
\label{eq:CLCsol1}
\Omega & =& \frac{\omega_1r_1^2 + \omega_2r_2^2}{r_1^2+r_2^2}, \quad\quad \sin(\Delta \phi) = -\frac{\gamma r_1r_2\Delta\omega}{k(r_1^2+r_2^2)}, \\
\label{eq:CLCsol2}
&&R-r_1^2 + m\Omega^2+k\left(\frac{r_2}{r_1}\cos(\Delta\phi)-1\right)=0, \\
\label{eq:CLCsol3}
&&R-r_2^2 + m\Omega^2+k\left(\frac{r_1}{r_2}\cos(\Delta\phi)-1\right)=0.
\end{eqnarray}
Equations~(\ref{eq:CLCsol2})-(\ref{eq:CLCsol3}) should be solved simultaneously to obtain $r_i$. For the CLCo regime it is non-trivial to obtain the analytic solution and reduces to finding the roots of a quartic equation but for the CLCi regime since $r_1=r_2=r$ we obtain $r^2 = R + m\Omega^2 + k[\cos(\Delta \phi)-1]$ with $\Omega = (\omega_1 + \omega_2)/2$ and $\cos(\Delta\phi) = \sqrt{1-\gamma^2\Delta\omega^2/4k^2}$. In the CLCi regime, the presence of an inertia-like term modifies only the radius ($r> 0$ since $4k^2 > \gamma^2\Delta\omega^2$ in the CLCi regime) of the limit cycle and not the frequency or phase difference. On the other hand in the CLCo regime the presence of an inertia-like term effects all parameters. 

In the large coupling $k$ limit, assuming that the radii of the two limit cycle oscillators are finite, Eqs.~(\ref{eq:CLCsol2}) and (\ref{eq:CLCsol3}) which need to be satisfied simultaneously reduce to
\begin{eqnarray}
\frac{r_2}{r_1}\cos(\Delta\phi) - 1 &=& 0,\nonumber \\
\frac{r_1}{r_2}\cos(\Delta\phi) - 1 &=& 0, \nonumber
\end{eqnarray}
which only permits one solution, i.e., the CLCi solution with $r_1 = r_2$, implying that the system would always end up coherently synchronized in phase for large couplings.\\ 
~\\
{\it Noisy regime .--}
In presence of noise, i.e., at a finite temperature, the noise averaged asymptotic solution decays to a fixed point as seen in Figs.~\ref{fig2} {\bf g} - {\bf j} (black and red solid lines). In this case, at the level of averages, we do not have a complex phase diagram and noise wipes out all signatures of the limit-cycle behavior and bistability in the asymptotic limit. The individual trajectories (grey lines in Figs.~\ref{fig2} {\bf g} - {\bf j}) carry huge fluctuations and may carry important correlations which can be wiped out in the noise averaged quantities like $\langle z_i \rangle$. Hence, these observables are not appropriate quantities to understand the behavior of the oscillators in different phases. In other words, the noisy ISLD does not exhibit different phases if we characterize its phase using the observables $\langle z_i \rangle$ as one would normally do in the deterministic case.

It is worth noting here that unlike the equilibrium scenario wherein phases can be fully characterized by the partition function, in nonequilibrium such a universal function does not exist~\cite{Kubo85}. Hence, one set of observables showing a lackluster behavior does not necessarily mean that \emph{all} observables will behave similarly. In the next section we will see how thermodynamic observables like heat and work retain information of the different phases and hence show universal behaviors depending on the phase of the oscillators.

\section{Stochastic Thermodynamics}
\label{sec:ST}
We are now interested in studying the dynamics of two coupled inertial-like Stuart-Landau oscillators in presence of stochastic forcing. Similar study was done in case of single Duffing oscillator using the Volterra series approach in \cite{BergerPRE19}, where the authors were mainly interested in developing an analytical formulation for the time series analysis of a simplest bistable system. We, however, want to understand the interplay of noise and synchronisation in our model. A non-linear system is often influenced due to its contact with an environment e.g. a thermal bath. Hence, studying the thermodynamic quantities like heat and work and their effect on the behaviour of the system becomes important. We are interested in understanding this interplay within the stochastic thermodynamics framework \cite{Sekimoto97,Seifert12,Murashita16}.

In order to build a thermodynamic picture of this model, we need to be able to identify meaningful physical quantities such as heat and work, as well as the energy of the system as a function of its coordinates. For an equilibrium system coupled with a single bath and an external forcing through a conservative interaction, those quantities would be straightforward to define: the energy could be identified in the logarithm of the equilibrium distribution of the system with constant external force, and subsequently the heat and work could be defined as the changes of energy of the system related, respectively, to stochastic transitions, and to external parameter variations.

Our first result in this section will be to show that, perhaps surprisingly, the ISLD does in fact fit in that framework, with each separate dimer being equivalent to a charged particle in a heat bath.

\subsection{Inertial-like SLO as a charged particle}

In this section, we show how a single noisy inertial-like SLO can be obtained from the dynamics of a charged particle subject to a quartic potential and a constant magnetic field, and connected to a heat bath. This allows us to justify the form of the parameters entering the SLO equation, and inform our thermodynamic analysis in the next section.

Consider an electric point particle with charge $q$ and mass $m$, with coordinates $X$ and $Y$, velocities $\dot X$ and $\dot Y$, subject to an external force deriving from a potential $U$ which we assume to be central in the plane $(X,Y)$ (e.g. the horizontal components of a potential electric field), and a Lorentz force due to a constant magnetic field $B$ normal to the plane $(X,Y)$. We also assume the particle is in contact with a heat bath (e.g. a viscous fluid) of strength $\gamma$ and temperature $T$, resulting in both a drag force and a random force. The equations of motion of this system are then
\begin{eqnarray}
m\ddot X&=&-\partial_XU+qB \dot Y-\gamma \dot X+\xi_X,\label{eq:EOMbareX}\\
m\ddot Y&=&-\partial_YU-qB \dot X-\gamma \dot Y+\xi_Y,\label{eq:EOMbareY}
\end{eqnarray}
where $\xi_X(t)$ and $\xi_Y(t)$ are Gaussian white noises with variance $\langle\xi_X(t)\xi_X(t')\rangle=\langle\xi_Y(t)\xi_Y(t')\rangle=2k_B T \gamma\delta(t-t')$. Note that the potential $U$ is arbitrary for now, to be as general as possible, but we will need to specify it to a quartic function later in order to recover the Stuart-Landau equations.

As is well known, the magnetic force does no work, though it is not quite a potential force. Let us define half the cyclotron frequency of the particle: $\omega=qB/2m$. The conservative part of the dynamics, obtained for $\gamma=0$, has then two conserved quantities: i) the energy
\begin{equation}
h=\frac{m}{2}(\dot X^2+\dot Y^2)+U(X^2+Y^2),
\end{equation}
and ii) the third component of the angular momentum
\begin{equation}
L_3=\vec{e}_3\cdot\left[\vec{r}\times(m\dot{\vec{r}}+q\vec{B}\times\vec{r})\right]=m\omega(X^2+Y^2)+m(X\dot Y-Y\dot X).
\end{equation}
Given the form of the coupling with the bath, which satisfies the Einstein relation with respect to the energy (and does not involve the angular momentum), the stationary distribution of this system is simply
\begin{equation}
\rho(X,Y,\dot X,\dot Y)=\mathcal{N}^{-1}\exp[-\beta h],\label{eq:StSLO}
\end{equation}
with the natural normalisation $\mathcal{N}=\int dX\int dY\int d\dot{X} \int d\dot{Y} \exp[-\beta h]$. Note that, even though $L_3$ is not relevant to our following analysis we give the relevant equations for the sake of completeness.

Defining $Z=X+iY$ and $\xi(t)=\xi^R(t)+i\xi^I(t)$, the equations of motion~(\ref{eq:EOMbareX}) and (\ref{eq:EOMbareY}) become
\begin{equation}\label{oscZ}
m\ddot Z=-\partial_{\bar{Z}} U(|Z|^2)-(i2m\omega+\gamma) \dot Z +\xi,
\end{equation}
with $\bar{Z}$ being the complex conjugate of $Z$. Let us define new coordinates $z=x+iy$ in the rotating frame through
\begin{equation}
z=\mathrm{e}^{i\omega t }Z.
\end{equation}
Placing ourselves in the rotating frame of angular velocity $\omega$ simplifies the velocity-dependent part of the dynamics, at the cost of an extra rotational force. Moreover, given that an isotopic Gaussian white noise is invariant under rotation, we deduce an equation for $\ddot z$:
\begin{equation}
m\ddot z=-\partial_{\bar{z}}\left(U(|z|^2)+m\omega^2|z|^2/2\right)+i\gamma\omega z-\gamma \dot z+\xi.
\end{equation}
If the potential $U$ is chosen to be quartic in $|z|$, we see that this becomes the noisy Stuart-Landau equation for a single oscillator. Importantly, the rotational term is proportional to $\gamma$, i.e., it is a virtual force resulting from friction in a rotating frame, and vanishes for $\gamma=0$, unlike the centrifugal force $-m\omega^2 z$. Note that, by construction, this system is in fact at equilibrium in a rotating frame, which does not affect its thermodynamics but gives its dynamics the appearance of non-equilibrium. All systems of this type are described by the so-called GENERIC formalism \cite{Grmela97, Grmela97II, Kraaij18}.

Let us therefore fix the value of the potential $U$ from here on. For the sake of simplicity, we will choose it such that it cancels the centrifugal term mentioned above:
\begin{equation}
 U(|z|^2)=\frac{|z|^4}{4}-(R+m\omega^2)\frac{|z|^2}{2}.
\end{equation}
We thus recover the uncoupled inertial-like SLO equation (\ref{ine_noise}) for $k=0$.

Under these new coordinates, the energy becomes
\begin{equation}\label{hSLO}
h=\frac{m}{2}\left[(\dot x+\omega y)^2+(\dot y-\omega x)^2\right]+U(x^2+y^2),
\end{equation}
giving us the corresponding stationary distribution, Eq.~(\ref{eq:StSLO}), and the third component of the angular momentum simplifies to
\begin{equation}
L_3=m(x\dot y-y\dot x).
\end{equation}
An important remark is that, although we have identified an energy (\ref{hSLO}) for the stationary distribution of a single SLO, this energy is not a Hamiltonian for the conservative part of the inertial-like Stuart-Landau equations, i.e. the equations of motion for $\gamma\rightarrow0$ do not derive from $h$ in the usual way. The reason for this is that the coordinate transformation from $Z$ to $z$ is not canonical, and this explains why the single SLO does not appear to be detail-balanced even though it is.

\subsection{Heat and work}
\label{sec:HeatnWork}
As we mentioned earlier, the Einstein relation is verified between the noise and the dissipation, relative to an inverse temperature $\beta$ and the Shannon entropy for independent particles. The heat exchanged with the reservoir is then unambiguously defined as the rate at which energy is lost by the system. Using the dynamics (\ref{oscZ}) of a single oscillator $i$ to compute this rate, and allowing for different values of the frequencies by denoting them $\omega_i$, we find
\begin{equation}
\delta Q_i=-\dot h_i|_{k=0}=\gamma(\dot X_i^2+\dot Y_i^2)-(\dot X_i\xi_i^R+\dot Y_i\xi_i^I)
\end{equation}
where the second part averages out to $0$ due to the noises. Since we dealt with only a single oscillator in the previous sub-section the quantities like energy $h$ or phase space variables $\{\dot x,x,\dot y,y\}$ did not have a sub-index but from now on the sub-index $i$ is the oscillator index. Note that both parts of the heat vanish for $\gamma\rightarrow0$, as expected. In terms of the rotating coordinates $z_i$, this becomes
\begin{eqnarray}
\delta Q_i&=&\gamma\left[(\dot x_i+\omega_i y_i)^2+(\dot y_i-\omega_i x_i)^2\right]
-\left[(\dot x_i+\omega_i y_i )\xi_i^R+(\dot y_i-\omega_i x_i)\xi_i^I\right].\nonumber
\end{eqnarray}
As a side note, the angular momentum exchanged by oscillator $i$ with the reservoir is given by
\begin{eqnarray}
-\dot{L}_3&=&m\gamma(X_i\dot Y_i-Y_i\dot X_i)+(X_i\xi_i^I-Y_i\xi_i^R)\\
&=&m\gamma(x_i\dot y_i-y_i\dot x_i)-m\gamma\omega(x_i^2+y_i^2)+(x_i\xi_i^I-y\xi_i^R).
\end{eqnarray}
where once again the noisy part averages out to $0$ and both terms disappear for $\gamma=0$.

Defining work is slightly less straightforward, as it requires us to make a meaningful choice of a reference potential meant to describe the system without external driving. Without a proper physical interpretation of the system and of the driving, any potential could be a candidate. In the present case, we have two interesting possibilities:\\
~\\
{\it Case I:} The first one is to consider the potential defined above for each oscillator separately, and define work as the energy variation in the system due to the inclusion of the interaction $k$. In this case, the undriven system is simply the two oscillators, each following Eq.~(\ref{oscZ}) with the appropriate frequency $\omega_i$. The driving consists in adding the terms proportional to $k$ in Eq.~(\ref{ine_noise}). The work rate can then be computed by applying this driven dynamics to the reference potential, in order to estimate the rate of energy change, and extracting the term which is not heat, which by construction is the term proportional to $k$. Hence, we get
\begin{eqnarray}\label{eq:Work1}
    \delta W^{(0)}&=&\dot h_1+\dot h_2+\delta Q_1+\delta Q_2 \\&=&-k\left[(x_1-x_2)(\dot x_1-\dot x_2)+(y_1-y_2)(\dot y_1-\dot y_2)+\Delta\omega(x_1 y_2-y_1 x_2)\right].\nonumber
\end{eqnarray}
~\\
{\it Case II:} Alternatively, we can consider the interaction term in Eq.~(\ref{ine_noise}) from a different reference system which is that of two charged particles coupled by a spring interaction $k$ in their original coordinate frame. This means adding a spring potential of strength $k$ in the original reference frame, so that the system has a total energy
\begin{eqnarray}
    H&=&h_1+h_2+\frac{k}{2}\left[(X_2-X_1)^2+(Y_2-Y_1)^2\right].
\end{eqnarray}
In the rotating frames, this extra term becomes more complicated because of the different frequencies:
\begin{eqnarray}
    H&=&h_1+h_2 +\frac{k}{2}\Big(x_1^2+y_1^2+x_2^2+y_2^2
    -2(x_1x_2+y_1y_2)\cos\left(\Delta\omega t\right)\nonumber\\
    &&-2(x_1y_2-y_1x_2)\sin\left(\Delta\omega t\right)\Big).
\end{eqnarray}
For the same reason, the interaction term in the reference equations of motion for $z_i$ is time-dependent in the rotating frame if the two frequencies $\omega_i$ are different, so that replacing it by a time-independent $k$ as seen in Eq.~(\ref{ine_noise}) amounts to having a work-performing external driving. In this case, when computing work in the same way as case I above, the first term of Eq.~(\ref{eq:Work1}) disappears, and we are left with
\begin{equation}
\label{eq:Work2}
\delta W^{(k)}=\dot H+\delta Q_1+\delta Q_2=-k\Delta\omega(x_1 y_2-y_1 x_2)
\end{equation}
which is surprisingly simple. Note that this case has the added advantage of producing no work in more situations ($\delta W^{(k)}=0$ for $\omega_1=\omega_2$ even if $k\neq 0$), whereas in case I this would lead to work that averages out to $0$ along any periodic trajectory. The definition of work we use is however ultimately a matter of choice, unless we have a physical reason to prefer one reference energy over the other. That being said, the difference between the two is a total time derivative, so that the choice does not affect the time-averaged output of the system.

\section{Numerical Analysis of Thermodynamic Work}\label{sec:Num}
We first analyze the zero temperature ($T=0$) noiseless limit. In this regime, both definitions of work [Eqs.~(\ref{eq:Work1}) and (\ref{eq:Work2})] for CLC and AD regions coincide with \begin{eqnarray}
\label{eq:Work3}
\delta W_{\rm AD} &=& 0, \nonumber \\
\label{eq:WorkCLC}
\delta W_{\rm CLC} &=& -\gamma\Delta\omega^2 \frac{r_1^2r_2^2}{r_1^2 + r_2^2}.
\end{eqnarray}
Above we have used $z_j=r_j e^{i(\Omega t + \phi_j)}$ with the parameters $r_j$ obtained using Eqs.~(\ref{eq:CLCsol2})-(\ref{eq:CLCsol3}). With the damping coefficient $\gamma > 0$ in the CLC regime the system always acts like a machine~\cite{Herpich18} with $\delta W < 0$ whereas no work is done in the AD regime since the system relaxes to a fixed point solution at the origin. Specifically, for the CLCi regime the average work can be computed analytically using $r_1=r_2=r$ provided below Eq.~(\ref{eq:CLCsol3}). For the general ILC regime wherein the amplitudes can be time dependent it is less clear whether the system works as a machine or a dissipator.

%%%%%%%%%% add (Jung-Wan) %%%%%%%%%
%%%%%%%%%% add (Jung-Wan) %%%%%%%%%
\begin{figure}
\begin{center}
\includegraphics[width=\columnwidth]{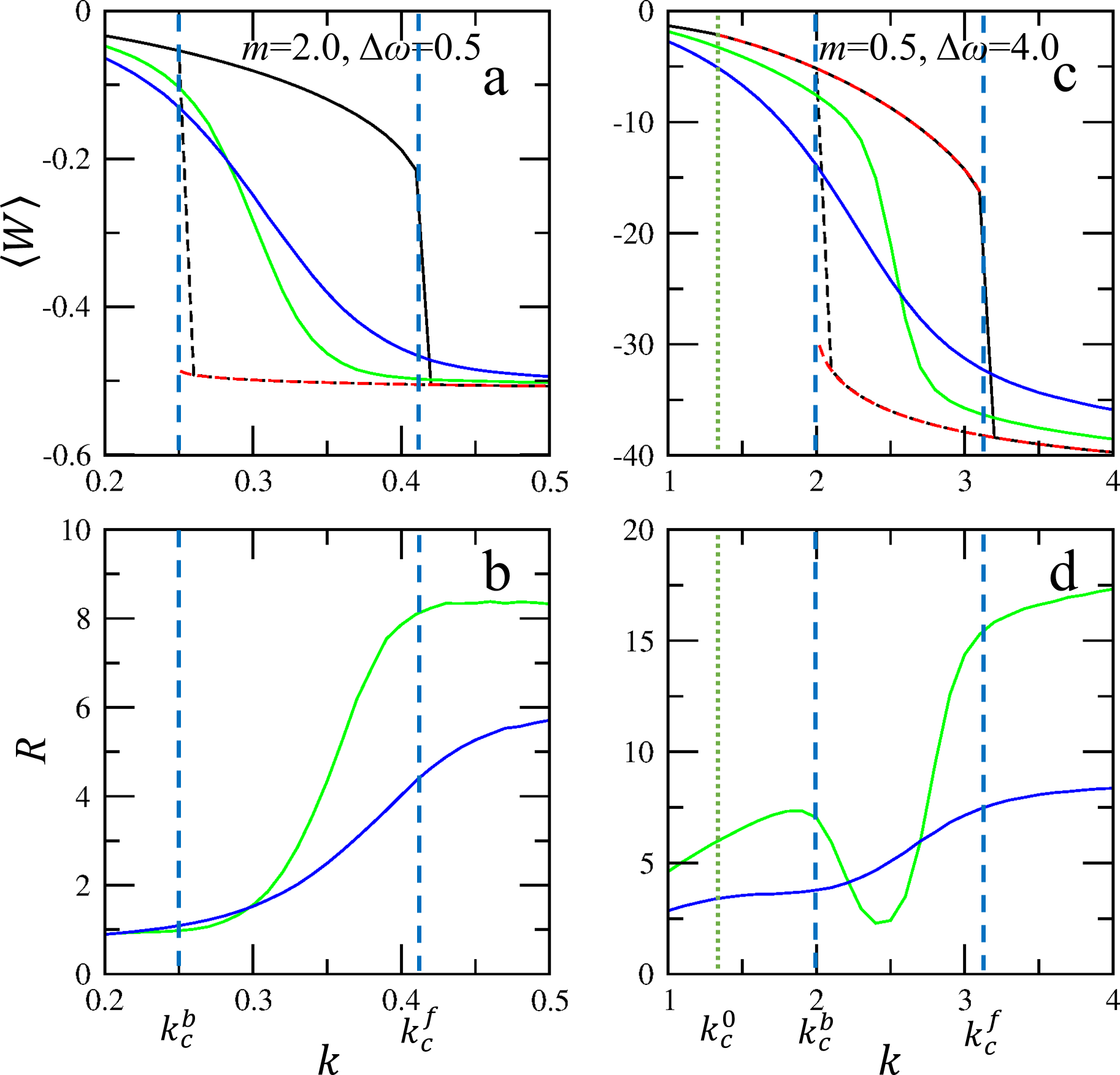}
\caption{Time average work $\langle W \rangle$, Eq.~(\ref{eq:Work1}), (panel {\bf a}) and reliability $R$, Eq.~(\ref{eq:reliability}), (panel {\bf b}) as a function of coupling constant $k$ with $k_{B}T_{1} = k_{B}T_{2} = 10^{-4}$ (black), $0.5$ (green), and $1.0$ (blue) when $m = 2.0$ and $\Delta \omega = 0.5$. Panels {\bf c} and {\bf d} correspond to $\langle W \rangle$ and $R$ for $k_{B}T_{1} = k_{B}T_{2} = 10^{-4}$ (black solid line), $1.0$ (green), and $4.0$ (blue) when $m = 0.5$ and $\Delta \omega = 4.0$. Solid lines correspond to forward direction initial conditions whereas the dashed lines correspond to the backward direction initial conditions, as determined by the noiseless system. For extremely low temperatures, wherein the dynamics are nearly deterministic, the reliability explodes due to the variance of work tending to zero. Hence, we do not plot reliability for $k_{B}T_{1} = k_{B}T_{2} = 10^{-4}$. At moderate to high temperatures (green and blue lines), the solutions from the forward (solid lines) and backward (dashed lines) direction initial conditions perfectly overlap. Red dashed line is the work obtained from analytic solutions, Eq.~(\ref{eq:Work3}), of CLCi and CLCo when $T=0$. The parameters used in panels {\bf a} and {\bf b} ({\bf c} and {\bf d}) are the same as Figs.~\ref{fig3} {\bf a} and {\bf b} ({\bf c} and {\bf d}) and the vertical lines denote the critical couplings $k$ where the system transits from one phase to another in the noiseless limit.}
\label{fig4}
\end{center}
\end{figure}
%%%%%%%%%%%%%%%
In the noisy regime, we integrate Eqs.~(\ref{Lx1}) - (\ref{Ly2}) numerically to study the thermodynamic response of the system. Simulations are performed using a modified velocity Verlet algorithm by Ermak \cite{Ermak78}. The algorithm was historically introduced to include hydrodynamic interactions in Brownian dynamics simulations, but for our purpose even the standard fourth order Runge-Kutta method would suffice. Thus, we now look at the average work $\langle W\rangle$ in different bifurcation regimes described in Fig.~\ref{fig3}. The system is first evolved for a transient time $ t = 20$ (see Figs.~\ref{fig2} {\bf g} - {\bf j} to gauge an estimate of the transient time). The average work at a specific $k$ is calculated after averaging over a duration of $8 \times 10^6$ with the incremental time step $dt \sim 10^{-3}$. The system is initiated in the same forward and reverse direction initial conditions we used for the noiseless case in Fig.~\ref{fig3}. The time averaged work $\langle W \rangle$ obtained from Eq.~(\ref{eq:Work1}) at different temperatures is shown in Fig.~\ref{fig4}. 

At extremely low temperatures (black lines) in Fig.~\ref{fig4} the simulation time is not long enough for the system to relax to its asymptotic state. Hence, in this case the system remains in a metastable state for extremely long times that mimics the noiseless system perfectly. In the metastable regime, the average work shows bistability with the forward (solid black lines) and backward (dashed black lines) initial condition results being distinctly different between the critical backward $k_c^b$ and critical forward $k_c^f$ couplings. In the CLC regime, the work output ($\langle W \rangle < 0$) of the machine matches perfectly with the analytic results obtained in the noiseless limit, Eq.~(\ref{eq:Work3}). 

\begin{figure}
\begin{center}
\includegraphics[width=\columnwidth]{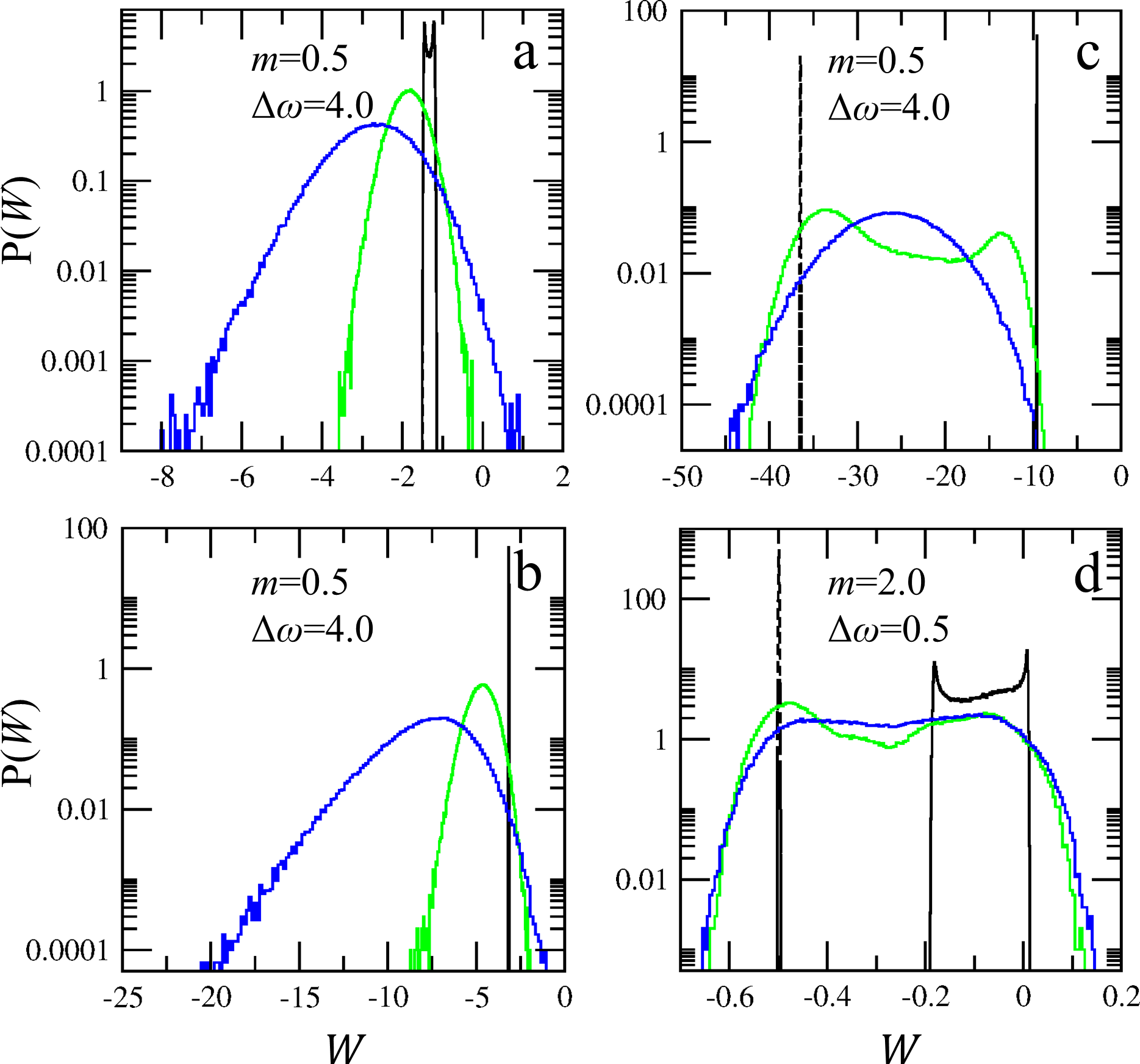}
\caption{PDF of the work $W$, Eq.~(\ref{eq:Work1}), for different values of $m$, $\Delta \omega$, and $k$, representative of different phases of the model. Panels {\bf a}, {\bf b}, and {\bf c} are with $m=0.5$, $\Delta \omega = 4.0$, and temperatures $k_{B}T_{1} = k_{B}T_{2} = 10^{-4}$ (black), $1.0$ (green), and $4.0$ (blue). Panel {\bf a} is within the ILC region with $k=1.0$, {\bf b} is for the CLCo region with $k=1.6$, and {\bf c} represents the bistable (CLCi + CLCo) region with $k=2.6$. Panel ${\bf d}$ is with $m=2.0$, $\Delta \omega = 0.5$, and temperatures $k_{B}T_{1} = k_{B}T_{2} = 10^{-4}$ (black), $0.5$ (green), and $1.0$ (blue) in the bistable (ILC + CLCi) region with $k=0.3$. Solid and dashed lines represent probability distributions of work $W$ obtained from the forward and backward initial conditions, respectively. Our low temperature results (black lines) are in the metastable regime in which the bistability is clearly reflected in the distinct distributions obtained for different initial conditions (panels {\bf c} and {\bf d}). For the green and blue lines, the solid and dashed lines perfectly overlap.}
\label{fig5}
\end{center}
\end{figure}

At moderate and high temperatures, we reach the asymptotic state in our simulation time and in this case the bistability disappears, i.e., the solid and dashed (blue and green) lines in Fig.~\ref{fig4} perfectly overlap. Thus, even though the system is in a unique steady state, the thermodynamic variable, work, retains critical information about the underlying phases obtained in the noiseless limit. Specifically, when the oscillators are incoherently synchronized the average work output is the minimum. Whereas, the magnitude of work output for the in-phase synchronized regime ($k > k_c^f$) is always larger than the incoherently or the out-of-phase coherently synchronized oscillators. Moreover, even the reliability~\cite{Talkner18, Son21},
\begin{eqnarray}
    R=\left| \frac{\langle W \rangle}{\sqrt{{\langle W^2 \rangle} - {\langle W \rangle}^2}} \right|
    \label{eq:reliability}
\end{eqnarray}
that measures how much the average dominates over the fluctuations, is the highest when the oscillators of the working substance synchronize in phase. In this phase, both work output and reliability decrease as a function of temperature indicating that the desirable machine is only obtained in the low temperature limit, concurrent with the notion of in-phase \emph{coherent} synchronization. The reliability of the machine is always larger when the oscillators are coherently synchronized. At moderate temperatures $k_BT_1=k_BT_2=1$, a sudden dip in the reliability is observed in the bistable II (CLCi + CLCo) regime around $k\approx 2.5$  (Fig.~\ref{fig4} {\bf d}). At these temperature values, the work distribution becomes bimodal with the position of the peaks being determined by the deterministic CLCi and CLCo work given by Eq.~(\ref{eq:WorkCLC}) [see description below on the distributions and Fig.~\ref{fig5} {\bf c}]. For the given choice of parameters, since the deterministic work values for CLCi and CLCo are well separated, the variance for the moderate temperature is large giving rise to a low reliability. Despite these occurrences that depend on the parameter choice, overall an in-phase synchronizing working substance always leads to a powerful and reliable machine. As a side remark, we would like to note that since we evaluate the time-averaged work output, our different definitions of work, Eqs.~(\ref{eq:Work1}) and~(\ref{eq:Work2}), match each other exactly, as expected.

Another question of interest, when looking at the inertial-like Stuart-Landau dimer as an engine, is that of efficiency. At first glance, given that the system works with a single temperature, using the first law of thermodynamics over a cycle (i.e. any trajectory of the system in its stationary or periodic regimes) implies that the heat absorbed is exactly converted to work, leading to unit efficiency. That being said, the concept of efficiency relies on a meaningful definition of \textit{absorbed} heat and of \textit{useful} work, which are unambiguous in the context of traditional heat engines, but less so for microscopic machines. We will therefore leave this question for more concrete situations, and simply note that, according to the noiseless expressions (\ref{eq:Work3}) that we obtained for the average work, this system operates on average as a useful machine in at least one regime, namely CLC.
%%%%%%%%%%%%%%%%%%%%%%%%%%%%%

%%%%%%%%%% add (Jung-Wan) %%%%%%%%%

The stochastic thermodynamics framework described in Sec.~\ref{sec:HeatnWork} allows us to go beyond the averages and calculate the full probability density function (PDF) of work as evaluated numerically in Fig.~\ref{fig5}. The system is evolved for a transient time ($t=20$) as described earlier and then the accumulated stochastic work is evaluated over a duration of period $\tau = 40$, which spans over multiple periods in the ILC regime for very low temperatures. This is repeated for $N=2\times 10^5$ times, for a total duration of $N\tau$, giving us the $N$ stochastic work values to form the distribution for particular values of the trap strength $k$ and temperatures. At small temperatures (black lines in Fig.~\ref{fig5}), our distributions are obtained in the metastable regime and can be fully explained using the noiseless solutions. In this regime, the noise only perturbs the work distribution slightly and gives it a finite width proportional to the strength of the noise, as expected. In cases wherein the CLC solutions hold, since the work (noiseless asymptotic limit) only depends on the \emph{time-independent} limit cycle radii $r_1$ and $r_2$, Eq.~(\ref{eq:WorkCLC}), the distributions are peaked at $W=\delta W_{\rm CLC} = -\gamma \Delta\omega^2 r_1^2r_2^2/(r_1^2 + r_2^2)$. For the ILC regimes, the broader distributions observed in panels {\bf a} and {\bf d} result from the fact that instantaneous work is not constant along the ILCs, so that the average work over a time-span $\tau$ will depend on the initial condition on the cycle. The range of values taken by $W$ will be of order $\tau^{-1}$ and will become negligible for $\tau$ large enough, though still noticeable if $\langle W\rangle$ is small, as is the case in panel {\bf d}.

At moderate temperatures (green curves in Fig.~\ref{fig5}), the asymptotic distributions are bimodal, which is a  signature of the underlying noiseless bistability. In the noiseless stable regimes, ILC and CLCo regime (Figs.~\ref{fig5} {\bf a} and {\bf b}) the work distributions broaden with temperature as expected, but in the bistable regimes, Figs.~\ref{fig5} {\bf c} and {\bf d}, the distributions become bimodal. In the bistable II regime (panel {\bf c}), wherein the two different limit cycles coexist (CLCi + CLCo), the system settles in the CLCi often indicated by the slightly higher maximum of the distribution centered around the noiseless CLCi distribution (dashed black line). The CLCo solution, which was stable in the noiseless limit, is no longer equally probable to the CLCi solution, but influences the distribution strongly making it bimodal. For the bistable I regime (panel {\bf d}), the system again tends to often be trapped in the CLCi solution. The ILC solution, which was stable in the noiseless limit, influences the overall distribution and leads to the overall distribution being bimodal. As temperature increases further (blue curves) all signatures of bistability are washed away and the variance of most observables including work scale with the temperature as can be observed from Fig.~\ref{fig5}.
%%%%%%%%%%%%%%%%%

\section{Conclusions}\label{sec:Con}
To conclude we studied the stochastic inertial-like Stuart-Landau dimer, whose coordinates $x_i$ and $y_i$ are independent of each other and represent the position of the oscillators. We extensively analysed our model in the zero temperature limit to uncover the rich phase space structure, which is not observed in the non-inertial system. We characterized various phases depending on the limit cycle behavior of the oscillators, namely, incoherent (ILC), in-phase coherent (CLCi), out-of-phase coherent (CLCo), and amplitude death (AD). Interestingly, we found the existence of bistable phases in which different limit cycles can coexist. Using a combination of analytic (linear stability analysis) and numeric (bifurcation diagrams) tools we obtained the phase boundaries that demarcate the different co-existing phases. At finite temperatures, the system always relaxed to a unique fixed point solution, erasing all information about the underlying phases. 

Furthermore, we developed a physically meaningful stochastic thermodynamics framework that required a frame transformation due to which the Stuart-Landau oscillator transformed to a magnetic charged particle in a quartic potential. Depending on the physical interpretation of the interaction between the two magnetic charged particles, stochastic heat and work were identified. More specifically, we considered two meaningful equilibrium systems of which the inertial-like Stuart-Landau dimer is then a driven version: one where the two charged particles are independent, and another where they interact through a spring potential. 

In the zero temperature regime, wherein the phase diagram displayed distinct phases analytic expressions for work were obtained in the CLC and AD regime. Surprisingly, the system output work like a machine in this regime and the behavior persisted for higher temperatures. In the ILC regime, although we were not able to prove this analytically, our extensive numerical simulations showed that the dimer continues to act as a machine albeit with a lower work output as compared to the in-phase coherently synchronized oscillators (CLCi). Other performance metrics, such as the reliability, corroborated that when the working substance is in-phase coherently synchronized we have the most desirable machine with a high reliability.

The signatures of the underlying bistability persisted for the averages in the temporally metastable regime at low temperatures. This led to a hysteresis curve for the average work output as the interaction between the oscillators was varied. Although such signatures were completely wiped out once the system reached the asymptotic state, but the distribution of work continued to display remnants of the underlying bistability. This was observed in the bimodal structure of the work distributions at moderate temperatures, but in the stable asymptotic limit. At extremely high temperatures, the thermal fluctuations dominated all processes and all signatures of bistability either in the averages or in the distributions disappeared. 

Our work highlights the importance of a synchronizing working substance in the performance of a machine. It not only shows the effect of phase transitions on thermodynamic observable, such as work, but also stresses that the stochastic thermodynamic framework need not be necessarily unique given the dynamical equations. In other words, stochastic thermodynamics should be a physically inspired and not a mathematically constructed framework using the equations of motion that govern the dynamics. Although, we considered a simple dimer in this study, extending our system to a network of oscillators to investigate the effect of long-range interactions~\cite{Gupta17}, the influence of network topology on thermodynamic observables, or to understand how biological systems make use of synchronization \cite{Hong20}, would be fascinating future directions.

\section*{Acknowledgements}
This research was supported by the Institute for Basic Science in Korea (IBS-R024-Y2 and IBS-R024-D1). J.T. would like to thank Peter Talkner for fruitful discussions.

\appendix
\section{Linear stability analysis}
\label{append:LSA}
In order to understand some of the phase boundaries shown in Fig.~\ref{fig2} we perform the linear stability analysis about the fixed point located at the origin. Rewriting Eq.~(\ref{ine_noise}), in the zero noise limit, using $v = \dot{z}$, we obtain,
\begin{eqnarray}
\label{Eqns_vel}
    \dot{z_1} &=& v_1, \\\nonumber
    m\dot{v_1} &=& -\gamma v_1 + (R + i \gamma\omega_1 -|z_1|^2)z_1 + k ( z_2 - z_1 ) + \xi_1 , \\\nonumber
    \dot{z_2} &=& v_2, \\\nonumber
    m\dot{v_2} &=& -\gamma v_2+ (R + i \gamma\omega_2 -|z_2|^2)z_2  + k ( z_1 - z_2 ) + \xi_2.
\end{eqnarray}
The linearized Jacobian Matrix $J$ in the zero noise limit at the origin, $z_j=0$ is given by
\begin{equation}
\label{Jacobian}
J = \frac{1}{m}\left(\begin{array}{cccc}
 0 & m & 0 & 0 \\
 R + i \gamma\omega_1 -k & -\gamma & k & 0 \\
 0 & 0 & 0 & m \\
 k & 0 & R + i \gamma\omega_2 -k & -\gamma 
\end{array}\right),
\end{equation}
where $\dot{Z} = J Z$ with $Z^T=(\delta z_1, \delta v_1, \delta z_2, \delta v_2)$ such that $\delta x$ are the small deviations about the origin. The eigenvalues $\lambda_i$ of $J$ are complex numbers. The real and imaginary parts are the decay (or growing) rates and the angular frequency of the orbit near the origin, respectively.

The stability of the fixed point, i.e., $z_j=0$, is determined by the maximal value of the real parts of complex eigenvalues. If the maximal value is negative, the origin is a stable fixed point. The boundaries of the AD region of Fig.~\ref{fig2} {\bf b} (orange lines), which is a Hopf bifurcation, can be obtained when the maximum real part of the eigenvalues of $J$ cross from positive to negative.

The four eigenvalues of the Jacobian matrix, Eq.~(\ref{Jacobian}), are given by
\begin{eqnarray}
\lambda_1 &=&-\frac{\gamma+\sqrt{\Omega_+}}{2m}, \quad \lambda_2=-\frac{\gamma-\sqrt{\Omega_+}}{2m}, \\
\lambda_3 &=&-\frac{\gamma+\sqrt{\Omega_-}}{2m}, \quad \lambda_4=-\frac{\gamma-\sqrt{\Omega_-}}{2m}, 
\end{eqnarray}
with $\Omega_{\pm} = \sqrt{\gamma^2+2m(2R-2k+i\gamma(\omega_1 +\omega_2) \pm \sqrt{(\gamma\Delta \omega)^2 -(2 k)^2})}$. At $\gamma\Delta\omega=\gamma(\omega_2-\omega_1)=2k$ the eigenvalues coalesce since $\Omega_+=\Omega_-$. This gives a degeneracy to the Jacobian eigenvalues causing this point in the phase diagram to be an exceptional line (red vertical lines in Figs.~\ref{fig2} {\bf a} - {\bf b}) that does not depend on the mass of the oscillators.


\begin{thebibliography}{100}
\bibitem{note1} The reverse direction indicates that the initial condition at any value of $k$ is chosen as the asymptotic limit of the coupling $k+\delta k$ along with a small perturbation.
\end{thebibliography}
\end{document}